\documentclass[preprint,preprintnumbers,amsmath,amssymb]{revtex4}

\usepackage{graphicx}
\usepackage{dcolumn}
\usepackage{bm}

\newcommand{\mb}[1]{ \mbox{\boldmath$#1$} }
\newcommand{\ds}{\displaystyle}
\newcommand{\beq}{\begin{eqnarray}}
\newcommand{\eeq}{\end{eqnarray}}
\newcommand{\beqq}{\begin{eqnarray*}}
\newcommand{\eeqq}{\end{eqnarray*}}
\newcommand{\p}{\partial}

\newcommand{\x}{\mbox{\boldmath$x$}}

\begin{document}

\title{BROWNIAN SIMULATIONS AND UNI-DIRECTIONAL FLUX IN DIFFUSION\\}
\author{A. Singer}
\email{amits@post.tau.ac.il}
\author{Z. Schuss}
\email{schuss@post.tau.ac.il} \affiliation{Department of Applied
Mathematics, Tel-Aviv University, Ramat-Aviv, 69978 Tel-Aviv,
Israel}

\begin{abstract}
\end{abstract}


\maketitle

\section{Introduction}
The prediction of ionic currents in protein channels of biological
membranes is one of the central problems of computational
molecular biophysics. None of the existing continuum descriptions
of ionic permeation captures the rich phenomenology of the patch
clamp experiments \cite{Hille}. It is therefore necessary to
resort to particle simulations of the permeation process
\cite{Im1}-\cite{Trudy1}. Computer simulations are necessarily
limited to a relatively small number of mobile ions, due to
computational difficulties. Thus a reasonable simulation can
describe only a small portion of the experimental setup of a patch
clamp experiment, the channel and its immediate surroundings. The
inclusion in the simulation of a part of the bath and its
connection to the surrounding bath are necessary, because the
conditions at the boundaries of the channel are unknown, while
they are measurable in the bath, away from the channel.

Thus the trajectories of the particles are described individually
for each particle inside the simulation volume, while outside the
simulation volume they can be described only by their statistical
properties. It follows that on one side of the interface between
the simulation and the surrounding bath the description of the
particles is discrete, while a continuum description has to be
used on the other side. This poses the fundamental question of how
to describe the particle trajectories at the interface, which is
the subject of this paper.

We address this problem for Brownian dynamics (BD) simulations,
connected to a practically infinite surrounding bath by an
interface that serves as both a source of particles that enter the
simulation and an absorbing boundary for particles that leave the
simulation. The interface is expected to reproduce the physical
conditions that actually exist on the boundary of the simulated
volume. These physical conditions are not merely the average
electrostatic potential and local concentrations at the boundary
of the volume, but also their fluctuation in time. It is important
to recover the correct fluctuation, because the stochastic
dynamics of ions in solution are nonlinear, due to the coupling
between the electrostatic field and the motion of the mobile
charges, so that averaged boundary conditions do not reproduce
correctly averaged nonlinear response. In a system of
noninteracting particles incorrect fluctuation on the boundary may
still produce the correct response outside a boundary layer in the
simulation region \cite{Spain}.

The boundary fluctuation consists of arrival of new particles from
the bath into the simulation and of the recirculation of particles
in and out of the simulation. The random motion of the mobile
charges brings about the fluctuation in both the concentrations
and the electrostatic field. Since the simulation is confined to
the volume inside the interface, the new and the recirculated
particles have to be fed into the simulation by a source that
imitates the influx across the interface. The interface does not
represent any physical device that feeds trajectories back into
the simulation, but is rather an imaginary wall, which the
physical trajectories of the diffusing particles cross and recross
any number of times. The efflux of simulated trajectories through
the interface is seen in the simulation, however, the influx of
new trajectories, which is the unidirectional flux (UF)  of
diffusion, has to be calculated so as to reproduce the physical
conditions, as mentioned above. Thus the UF is the source strength
of the influx, and also the number of trajectories that cross the
interface in one direction per unit time.

The mathematical problem of the UF begins with the description of
diffusion by the diffusion equation. The diffusion equation (DE)
is often considered to be an approximation of the Fokker-Planck
equation (FPE) in the Smoluchowski limit of large damping. Both
equations can be written as the conservation law
 \beq
 \frac{\p p}{\p t}=-\nabla\cdot\mb{J}.\label{CL}
 \eeq
The flux density $\mb{J}$ in the diffusion equation is given by
 \beq
 \mb{J}(\x,t)=-\frac1\gamma\left[\varepsilon\nabla p(\x,t)-\mb{f}(\x)p(\x,t)\right],
 \label{Jdiff}
 \eeq
where $\gamma$ is the friction coefficient (or dynamical
viscosity), $\varepsilon=\ds\frac{k_BT}{m}$, $k_B$ is Boltzmann's
constant, $T$ is absolute temperature, and $m$ is the mass of the
diffusing particle. The external acceleration field is
$\mb{f}(\x)$ and $p(\x,t)$ is the density (or probability density)
of the particles \cite{book}. The flux density in the FPE is given
by where the net probability flux density vector has the
components
 \beq
 J_{\x}(\x,\mb{v},t)&=&\mb{v}p(\x,\mb{v},t)\nonumber\\
 &&\label{FPflux}\\
 J_{\mb{v}}(\x,\mb{v},t)&=&-\left(\gamma
 \mb{v}-\mb{f}(\x)\right)p(\x,\mb{v},t) -\varepsilon\gamma\nabla_{\mb{v}}p(\x,\mb{v},t).
\nonumber
 \eeq

The density $p(\x,t)$ in the diffusion equation (\ref{CL}) with
(\ref{Jdiff}) is the probability density of the trajectories of
the Smoluchowski stochastic differential equation
 \beq
\dot{\x}=\frac1\gamma\mb{f}(\x)+\sqrt{\frac{2\varepsilon}\gamma}\,\dot{\mb{w}},\label{SD}
 \eeq
where $\mb{w}(t)$ is a vector of independent standard Wiener
processes (Brownian motions).

The density $p(\x,\mb{v},t)$ is the probability density of the
phase space trajectories of the Langevin equation
 \beq
 \ddot{\x}+\gamma\dot{\x}=\mb{f}(\x)+\sqrt{2\varepsilon\gamma}\dot{\mb{w}}.\label{LD}
 \eeq

In practically all conservation laws of the type (\ref{CL})
$\mb{J}$ is a {\em net flux density} vector. It is often necessary
to split it into two unidirectional components across a given
surface, such that the net flux  $\mb{J}$ is their difference.
Such splitting is pretty obvious in the FPE, because the velocity
$\mb{v}$ at each point $\x$ tells the two UFs apart. Thus, in one
dimension,
 \beq
 J_{LR}(x,t)&=&\int_0^\infty vp(x,v,t)\,dv,\quad J_{RL}(x,t)=-\int_{-\infty}^0
 vp(x,v,t)\,dv\nonumber\\
 &&\label{Js}\\
J_{\mbox{net}}(x,t)&=&J_{LR}(x,t)-J_{RL}(x,t)=\int_{-\infty}^\infty
vp(x,v,t)\,dv.\nonumber
 \eeq

In contrast, the net flux $J(x,t)$ in the DE cannot be split this
way, because velocity is not a state variable. Actually, the
trajectories of a diffusion process do not have well defined
velocities, because they are nowhere differentiable with
probability 1 \cite{DEK}. These trajectories cross and recross
every point $x$ infinitely many times in any time interval
$[t,t+\Delta t]$, giving rise to infinite UFs. However, the net
diffusion flux is finite, as indicated in eq.(\ref{Jdiff}). This
phenomenon was discussed in detail in \cite{unidirect}, where a
path integral description of diffusion was used to define the UF.
The unidirectional diffusion flux, however, is finite at absorbing
boundaries, where the UF equals the net flux. The UFs measured in
diffusion across biological membranes by using radioactive tracer
\cite{Hille} are in effect UFs at absorbing boundaries, because
the tracer is a separate ionic species \cite{BobDuan}.

An apparent paradox arises in the Smoluchowski approximation of
the FPE by the DE, namely, the UF of the DE is infinite for all
$\gamma$, while the UF of the FPE remains finite, even in the
limit $\gamma\to\infty$, in which the solution of the DE is an
approximation of that of the FPE \cite{EKS}. The ``paradox" is
resolved by a new derivation of the FPE for LD from the Wiener
path integral. This derivation is different than the derivation of
the DE or the Smoluchowski equation from the Wiener integral (see,
e.g. \cite{Kleinert}-\cite{Freidlin}) by the method of M. Kac
\cite{Kac}. The new derivation shows that the path integral
definition of UF in diffusion, as first introduced in
\cite{unidirect}, is consistent with that of UF in the FPE.
However, the definition of flux involves the limit $\Delta t\to0$,
that is, a time scale shorter than $1/\gamma$, on which the
solution of the DE is not a valid approximation to that of the
FPE.

This discrepancy between the Einstein and the Langevin
descriptions of the random motion of diffusing particles was
hinted at by both Einstein and Smoluchowski. Einstein
\cite{Einstein} remarked that his diffusion theory is based on the
assumption that the diffusing particles are observed
intermittently at short time intervals, but not too short, while
Smoluchowski \cite{Smoluchowski06} showed that the variance of the
displacement of Langevin trajectories is quadratic in $t$ for
times much shorter than the relaxation time $1/\gamma$, but is
linear in $t$ for times much longer that $1/\gamma$, which is the
same as in Einstein's theory of diffusion \cite{Chandra}.

The infinite unidirectional diffusion flux imposes serious
limitations on BD simulations of diffusion in a finite volume
embedded in a much larger bath. Such simulations are used, for
example, in simulations of ion permeation in protein channels of
biological membranes \cite{Hille}. If parts of the bathing solutions
on both sides of the membrane are to be included in the simulation,
the UFs of particles into the simulation have to be calculated.
Simulations with BD would lead to increasing influxes as the time
step is refined.

The method of resolution of the said ``paradox" is based on the
definition of the UF of the Langevin dynamics (LD) in terms of the
Wiener path integral, analogous to its definition for the BD in
\cite{unidirect}. The UF $J_{LR}(x,t)$ is the probability per unit
time $\Delta t$ of trajectories that are on the left of $x$ at
time $t$ and are on the right of $x$ at time $t+\Delta t$. We show
that the UF of BD coincides with that of LD if the time unit
$\Delta t$ in the definition of the unidirectional diffusion flux
is exactly
 \beq
 \Delta t=\frac2\gamma.\label{2g}
 \eeq
We find the strength of the source that ensures that a given
concentration is maintained on the average at the interface in a
BD simulation. The strength of the left source $J_{LR}$ is to
leading order independent of the efflux and depends on the
concentration $C_L$, the damping coefficient $\gamma$, the
temperature $\varepsilon$, and the time step $\Delta t$, as given
in eq.(\ref{JLRexp}). To leading order it is
 \beq
 J_{LR}=\sqrt{\frac\varepsilon{\pi\gamma\Delta
 t}}\,C_L+O\left(\frac1\gamma\right).\label{JLRi}
 \eeq

We also show that the coordinate of a newly injected particle has
the probability distribution of the residual of the normal
distribution. Our simulation results show that no spurious
boundary layers are formed with this scheme, while they are formed
if new particles are injected at the boundary. The simulations
also show that if the injection rate is fixed, there is depletion
of the population as the time step is refined, but there is no
depletion if the rate is changed according to eq.(\ref{JLRi}).

In Section \ref{sec:derivation}, we derive the FPE for the LD
(\ref{LD}) from the Wiener path integral. In Section
\ref{sec:unidir-langevin}, we define the unidirectional
probability flux for LD by the path integral and show that is
indeed given by (\ref{Js}). In Section \ref{sec:eks}, we use the
results of \cite{EKS} to calculate explicitly the UF in the
Smoluchowski approximation to the solution of the FPE and to
recover the flux (\ref{Jdiff}). In Section \ref{sec:unidir-smol},
we use the results of \cite{unidirect} to evaluate the UF of the
BD trajectories (\ref{SD}) in a finite time unit $\Delta t$. In
the limit $\Delta t\to0$ the UF diverges, but if it is chosen as
in (\ref{2g}), the UFs of LD and BD coincide. In Section
\ref{sec:simulation} we describe the a BD simulation of diffusion
between fixed concentrations and give results of simulations.
Finally, Section \ref{sec:summary} is a summary and discussion of
the results.

\section{Derivation of the Fokker-Planck equation
from a path integral} \label{sec:derivation}

The LD (\ref{LD}) of a diffusing particle can be written as the
phase space system
 \beq
\dot x=v,\quad \dot v=-\gamma
v+f(x)+\sqrt{2\varepsilon\gamma}\,\dot w.\label{Lang}
 \eeq
This means that in time $\Delta t$ the dynamics progresses
according to
 \beq
x(t+\Delta t)&=&x(t)+v(t)\Delta t+o(\Delta t)\label{Delta1}\\
&&\nonumber\\
v(t+\Delta t)&=&v(t)+[-\gamma v(t)+f(x(t))]\Delta
t+\sqrt{2\varepsilon\gamma}\,\Delta w+o(\Delta t),\label{Delta2}
 \eeq
where $\Delta w\sim{\cal N}(0,\Delta t)$, that is, $\Delta w$ is
normally distributed with mean 0 and variance $\Delta t$. This
means that the probability density function evolves according to
the propagator
 \beq
 &&\mbox{Prob}\{x(t+\Delta t)=x,v(t+\Delta t)=v\}=p(x,v,t+\Delta t)=o(\Delta t)+\label{propagator}\\
 &&\nonumber\\
 &&\frac1{\sqrt{4\varepsilon\gamma\pi\Delta
 t}}\int_{a}^{b}\!\int_{-\infty}^{\infty}
 p(\xi,\eta,t)\delta(x-\xi-\eta\Delta
 t)\exp\left\{-\frac{\left[v-\eta-[-\gamma\eta+f(\xi)]\Delta
 t\right]^2}{4\varepsilon\gamma\Delta t}\right\}\,d\xi\,d\eta.\nonumber
 \eeq
To understand (\ref{propagator}), we note that given that the
displacement and velocity of the trajectory at time $t$ are
$x(t)=\xi$ and $v(t)=\eta$, respectively, then according to
eq.(\ref{Delta1}), the displacement of the particle at time
$t+\Delta t$ is deterministic, independent of the value of $\Delta
w$, and is $x=\xi + \eta \Delta t+o(\Delta t)$. Thus the
probability density function (pdf) of the displacement is
$\delta(x-\xi-\eta \Delta t+o(\Delta t))$. It follows that the
displacement contributes to the joint propagator
(\ref{propagator}) of $x(t)$ and $v(t)$ a multiplicative factor of
the Dirac $\delta$ function. Similarly, eq.(\ref{Delta2}) means
that the conditional pdf of the velocity at time $t+\Delta t$,
given $x(t)=\xi$ and $v(t)=\eta$, is normal with mean
$\eta+[-\gamma \eta + f(\xi)]\Delta t+o(\Delta t)$ and variance
$2\epsilon\gamma\Delta t+o(\Delta t)$, as reflected in the
exponential factor of the propagator. If trajectories are
terminated at the ends of an finite or infinite interval $(a,b)$,
the integration over the displacement variable extends only to
that interval.

The basis for our analysis of the UF is the following new
derivation of the Fokker-Planck equation from
eq.(\ref{propagator}). Integration with respect to $\xi$ gives
 \beq
 &&p(x,v,t+\Delta t)=o(\Delta t)+\label{propagator1}\\
 &&\nonumber\\
 &&\frac1{\sqrt{4\varepsilon\gamma\pi\Delta
 t}}\int_{-\infty}^\infty
 p(x-\eta\Delta t,\eta,t)
\exp\left\{-\frac{\left[v-\eta-[-\gamma\eta+f(x-\eta\Delta
t)]\Delta t\right]^2}{4\varepsilon\gamma\Delta
t}\right\}\,d\eta.\nonumber
 \eeq
Changing variables to
 \[-u=\frac{v-\eta-[-\gamma\eta+f(x-\eta\Delta
 t)]\Delta t}{{\sqrt{2\varepsilon\gamma\Delta t}}},\]
and expanding in powers of $\Delta t$, the integral takes the form
 \beq
 &&p(x,v,t+\Delta t)=\frac1{\sqrt{2\pi}(1-\gamma\Delta t +
 o(\Delta t))}\int_{-\infty}^\infty \,e^{-u^2/2}\,du\times\label{pxyt}\\
 &&\nonumber\\
 && p(x-v(1+\gamma\Delta t)\Delta t+o(\Delta t),v(1+\gamma\Delta t)+u\sqrt{2\varepsilon\gamma\Delta
 t}-f(x)\Delta t(1+\gamma\Delta t)+o(\Delta t),t)\nonumber
 \eeq
Reexpanding in powers of $\Delta t$, we get
 \beqq
&& p(x-v(1+\gamma\Delta t)\Delta t+o(\Delta t),v(1+\gamma\Delta
t)+u\sqrt{2\varepsilon\gamma\Delta
 t}-f(x)\Delta t(1+\gamma\Delta t)+o(\Delta t),t)=\\
 &&\\
 &&p(x,v,t)-v\Delta t\frac{\p p(x,v,t)}{\p x}+\frac{\p
 p(x,v,t)}{\p v}\left(v\gamma\Delta
 t+u\sqrt{2\varepsilon\gamma\Delta t}-f(x)\Delta
 t+o(\Delta t)\right)+\\
 &&\\
 &&\varepsilon\gamma u^2\Delta t\frac{\p^2p(x,v,t)}{\p
 v^2}+o(\Delta t),
 \eeqq
so (\ref{pxyt}) gives
 \beq
p(x,v,t+\Delta t)-\frac{p(x,v,t)}{1-\gamma\Delta t}&=&-
 \frac1{1-\gamma\Delta t}v\Delta t\frac{\p p(x,v,t)}{\p x}+\frac{\Delta
 t}{1-\gamma\Delta t}\frac{\p
 p(x,v,t)}{\p v}\left(v\gamma-f(x)\right)\nonumber\\
 &&\nonumber\\
 &+&\frac{\varepsilon\gamma\Delta t}{1-\gamma\Delta t}\frac{\p^2p(x,v,t)}{\p
 v^2}+O\left(\Delta t^{3/2}\right).\nonumber
 \eeq
Dividing by $\Delta t$ and taking the limit $\Delta t\to0$, we
obtain the Fokker-Planck equation in the form
 \beq
 \frac{\p p(x,v,t)}{\p t}=-v\frac{\p p(x,v,t)}{\p x}+\frac\p{\p v}\left[\left(\gamma
 v-f(x)\right)p(x,v,t)\right] +\varepsilon\gamma\frac{\p^2p(x,v,t)}{\p
 v^2},\label{FPE}
 \eeq
which is the conservation law (\ref{CL}) with the flux components
(\ref{FPflux}). The UF $J_{LR}(x_1,t)$ is usually defined as the
integral of $J_x(x_1,v,t)$ over the positive velocities \cite[and
references therein]{EKS}, that is,
 \beq
J_{LR}(x_1,t)=\int_0^\infty vp(x_1,v,t)\,dv.\label{JLRFPE}
 \eeq
To show that this integral actually represents the probability of
the trajectories that move from left to right across $x_1$ per
unit time, we evaluate below the probability flux from a path
integral.

\section{The unidirectional flux of the Langevin equation}
\label{sec:unidir-langevin}

The instantaneous unidirectional probability flux from left to
right at a point $x_1$ is defined as the probability per unit time
($\Delta t$), of Langevin trajectories that are to the left of
$x_1$ at time $t$ (with any velocity) and propagate to the right
of $x_1$ at time $t+\Delta t$ (with any velocity), in the limit
$\Delta t\to0$. This can be expressed in terms of a path integral
on Langevin trajectories on the real line as
 \beq
J_{LR}(x_1,t)&=&\lim_{\Delta t\to0}\frac1{\Delta
t}\int_{-\infty}^{x_1}\,d\xi\int_{x_1}^\infty\,dx\int_{-\infty}^\infty\,d\eta
\int_{-\infty}^\infty\,dv\
\frac1{\sqrt{4\varepsilon\gamma\pi\Delta t}}p(\xi,\eta,t)
 \delta(x-\xi-\eta\Delta
 t)\times\nonumber\\
 &&\nonumber\\
 &&\exp\left\{-\frac{\left[v-\eta-[-\gamma\eta+f(\xi)]\Delta
 t\right]^2}{4\varepsilon\gamma\Delta t}\right\}.\label{JLRDEF}
 \eeq
Integrating with respect to $v$ eliminates the exponential factor
and integration with respect to $\xi$ fixes $\xi$ at $x-\eta\Delta
t$, so
 \beq
J_{LR}(x_1,t)&=&\lim_{\Delta t\to0}\frac1{\Delta
t}\int\!\int_{x-\eta\Delta t<x_1}
  p(x-\eta\Delta t,\eta,t)\,d\eta\,dx\nonumber\\
&&\nonumber\\
&=&\lim_{\Delta t\to0}\frac1{\Delta
t}\int_0^\infty\,d\eta\int_{x_1-\eta\Delta
t}^{x_1}p(u,\eta,t)\,du\nonumber\\
&&\nonumber\\
&=&\int_0^\infty\eta p(x_1,\eta,t)\,d\eta.\label{JLR}
 \eeq
The expression (\ref{JLR}) is identical to (\ref{JLRFPE}).

\section{The Smoluchowski approximation to the
unidirectional current}\label{sec:eks}

The following calculations were done in \cite{EKS} and are shown
here for completeness. In the overdamped regime, as
$\gamma\to\infty$, the Smoluchowski approximation to $p(x,v,t)$ is
given by
 \beq
&&p(x,v,t)\sim\frac{e^{-v^2/2\epsilon}}{\sqrt{2\pi
\epsilon}}\left\{p(x,t) -\frac{1}{\gamma}\left[\frac{\partial
p(x,t)}{\partial x}-\frac{1}{\epsilon}f(x)p(x,t)\right]v
+O\left(\frac{1}{\gamma^2}\right)\right\},\label{expansion}
 \eeq
where the marginal density $p(x,t)$ satisfies the
Fokker-Planck-Smoluchowski equation
\begin{eqnarray}
\gamma\frac{\partial p(x,t)}{\partial t}=\varepsilon
\frac{\partial^2p(x,t)}{\partial x^2}-\frac{\partial}{
\partial x}\left[f(x) p(x,t)\right].  \label{FPEQ}
\end{eqnarray}

According to (\ref{JLRFPE}) and (\ref{expansion}), the UF is
 \beq
J_{LR}(x_1,t)&=&\int_0^\infty vp(x_1,v,t)\,dv\nonumber\\
&&\nonumber\\
&=&\int_0^\infty v \frac{e^{-v^2/2\epsilon}}{\sqrt{2\pi
\epsilon}}\left\{p(x,t) -\frac{1}{\gamma}\left[\frac{\partial
p(x,t)}{\partial x}-\frac{1}{\epsilon}f(x)p(x,t)\right]v
+O\left(\frac{1}{\gamma^2}\right)\right\}\,dv \nonumber \\
&&\nonumber \\
&=& \sqrt{\frac\varepsilon{2\pi}}
p(x_1,t)-\frac{1}{2\gamma}\left[\varepsilon\frac{\partial
p(x,t)}{\partial x}-f(x)p(x,t)\right]
+O\left(\frac{1}{\gamma^2}\right).\label{unismolLR}
 \eeq
Similarly, the UF from right to left is
 \beq
J_{RL}(x_1,t)&=&-\int^0_{-\infty} vp(x_1,v,t)\,dv\nonumber\\
&&\nonumber\\
&=& \sqrt{\frac\varepsilon{2\pi}}
p(x_1,t)+\frac{1}{2\gamma}\left[\varepsilon\frac{\partial
p(x,t)}{\partial x}-f(x)p(x,t)\right]
+O\left(\frac{1}{\gamma^2}\right).\label{unismolRL}
 \eeq
Both UFs in (\ref{unismolLR}) and (\ref{unismolRL}) are finite and
proportional to the marginal density at $x_1$. The net flux is the
difference
 \beq
 J_{\mbox{net}}(x_1,t)=J_{LR}(x_1,t)-J_{RL}(x_1,t)=
 -\frac{1}{\gamma}\left[\varepsilon\frac{\partial
p(x,t)}{\partial x}-f(x)p(x,t)\right],\label{JNETSmol}
 \eeq
as in classical diffusion theory \cite{EKS}, \cite{Gardiner}.

\section{The unidirectional current in the
Smoluchowski equation} \label{sec:unidir-smol}

Classical diffusion theory, however, gives a different result. In
the overdamped regime the Langevin equation (\ref{Lang}) is
reduced to the Smoluchowski equation \cite{book}
 \beq
 \gamma\dot x=f(x)+\sqrt{2\varepsilon\gamma}\,\dot w.\label{Smol}
 \eeq
As in Section \ref{sec:unidir-langevin}, the unidirectional
probability current (flux) density at a point $x_1$ can be expressed
in terms of a path integral as
\begin{equation}
J_{LR}(x_1,t)=\lim_{\Delta t\rightarrow 0}J_{LR}(x_1,t,\Delta t),
\label{JLRN}
\end{equation}
where
\begin{eqnarray}
&&J_{LR}(x_1,t,\Delta t)=\sqrt{\frac{ \gamma}{4\pi\varepsilon
\Delta t}}\int_0^\infty d\xi \int_\xi ^\infty d\zeta\,\exp \left\{
-\frac{\gamma\zeta ^2}{4\varepsilon}\right\}\times\label{JLRexp}
\\
&&\nonumber\\
&&\left\{ p\left( x_1,t\right) -\sqrt{\Delta t}\left[-\frac{\zeta
f(x_1)}{2\varepsilon} p\left( x_1,t\right) +\left( \zeta -\xi
\right) \frac{\p p(x_1,t)}{\p x} \right] +O\left( \frac{\Delta
t}\gamma\right) \right\} . \nonumber
\end{eqnarray}

It was shown in \cite{unidirect} that
\begin{eqnarray}
&&J_{LR}(x_1,t,\Delta t)=\sqrt{\frac\varepsilon{\pi\gamma\Delta
t}}\,p(x_1,t)+\frac1{2\gamma}\left( f(x_1)p(x_1,t)
-\varepsilon\frac{\p p(x_1,t)}{\p
x}\right)+O\left(\frac{\sqrt{\Delta
t}}{\gamma^{3/2}}\right).\label{JLRexp}
\end{eqnarray}
Similarly,
\[ J_{RL}(x_{1},t)=\lim_{\Delta t\rightarrow 0}J_{RL}(x_{1},t,\Delta t),\]
where
\begin{eqnarray}
&&J_{RL}(x_1,t,\Delta t)=\sqrt{\frac{ \gamma}{4\pi\varepsilon
\Delta t}}\int_0^\infty d\xi \int_\xi ^\infty d\zeta\,\exp \left\{
-\frac{\gamma\zeta ^2}{4\varepsilon}\right\}\times\nonumber\\
&&\nonumber\\
&&\left\{ p\left( x_1,t\right) +\sqrt{\Delta t}\left[-\frac{\zeta
f(x_1)}{2\varepsilon} p\left( x_1,t\right) +\left( \zeta -\xi
\right)\frac{\p p(x_1,t)}{\p x} \right] +O\left( \frac{\Delta
t}\gamma\right) \right\} .
\nonumber\\
&&\nonumber\\
&=&\sqrt{\frac\varepsilon{\pi\gamma\Delta
t}}\,p(x_1,t)-\frac1{2\gamma}\left( f(x_1)p(x_1,t)
-\varepsilon\frac{\p p(x_1,t)}{\p
x}\right)+O\left(\frac{\sqrt{\Delta
t}}{\gamma^{3/2}}\right).\label{JRLexp}
\end{eqnarray}

If $p(x_{1},t)>0$, then both $J_{LR}(x_{1},t)$ and
$J_{RL}(x_{1},t)$ are infinite, in contradiction to the results
(\ref{unismolLR}) and (\ref{unismolRL}). However, the net flux
density is finite and is given by
\begin{eqnarray}
J_{\mbox{net}}(x_{1},t)&=&\lim_{\Delta t\rightarrow0}
\left\{J_{LR}(x_{1},t,\Delta t)-J_{RL}(x_{1},t,\Delta
t)\right\}\nonumber\\
&&\nonumber\\
&=&-\frac1\gamma\left[\varepsilon\frac{\partial}{\partial x}
p(x_1,t)-{f(x_1)}p(x_1,t)\right], \label{JNET}
\end{eqnarray}
which is identical to (\ref{JNETSmol}).

The apparent paradox is due to the idealized properties of the
Brownian motion. More specifically, the trajectories of the
Brownian motion, and therefore also the trajectories of the
solution of eq.(\ref{Smol}), are nowhere differentiable, so that
any trajectory of the Brownian motion crosses and recrosses the
point $x_1$ infinitely many times in any time interval
$[t,t+\Delta t]$ \cite{ItoMcKean}. Obviously, such a vacillation
creates infinite UFs.

Not so for the trajectories of the Langevin equation (\ref{Lang}).
They have finite continuous velocities, so that the number of
crossing and recrossing is finite. We note that setting
$\gamma\Delta t=2$ in equations (\ref{JLRexp}) and (\ref{JRLexp})
gives (\ref{unismolLR}) and (\ref{unismolRL}).

\section{Brownian Simulations}
\label{sec:simulation} 
\begin{figure}
\includegraphics[width=7in]{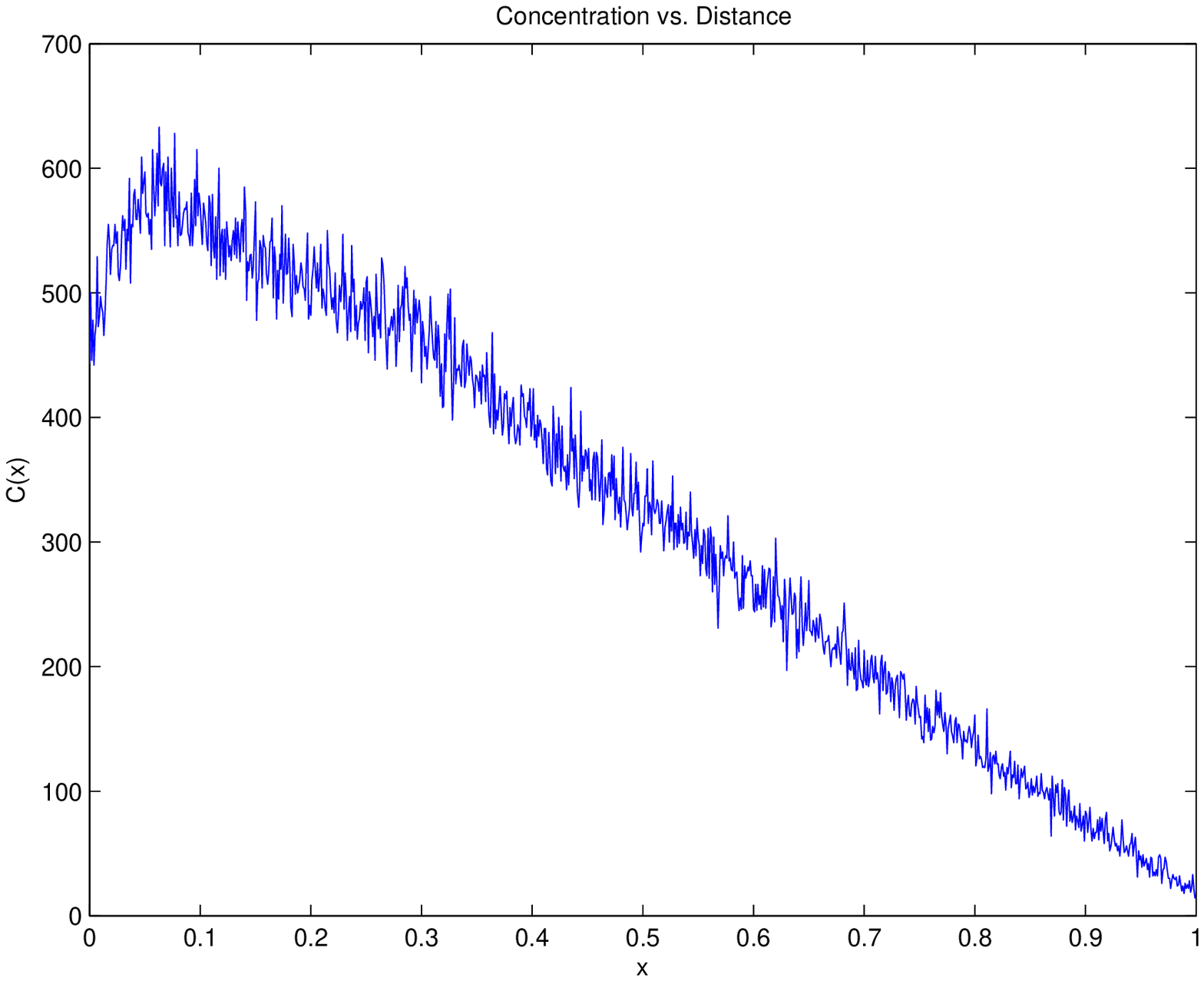}
\caption{The concentration profile of Brownian trajectories that
are initiated at $x=0$ with a normal distribution, and terminated
at either $x=0$ or $x=1$. \label{f:bump}}
\end{figure}

Here we design and analyze a BD simulation
of particles diffusing between fixed concentrations. Thus, we
consider the free Brownian motion (i.e., $\mb{f}=0$ in eq.
(\ref{SD})) in the interval $[0,1]$. The trajectories were
produced as follows: a) According to the dynamics (\ref{SD}), new
trajectories that are started at $x(-\Delta t)=0$ move to
$x(0)=\sqrt{\ds\frac{2\varepsilon}{\gamma}}\left|\Delta w\right|$;
b) The dynamics progresses according to the Euler scheme
$x(t+\Delta t) = x(t) +
\sqrt{\ds\frac{2\varepsilon}{\gamma}}\Delta w$; c) The trajectory
is terminated if $x(t)>1$ or $x(t)<0$. The parameters are
$\varepsilon=1,\,\gamma=1000,\,\Delta t=1$. We ran 10,000 random
trajectories and constructed the concentration profile by dividing
the interval into equal parts and recording the time each
trajectory spent in each bin prior to termination. The results are
shown in Figure \ref{f:bump}. The simulated concentration profile
is linear, but for a small depletion layer near the left boundary
$x=0$, where new particles are injected. This is inconsistent with
the steady state DE, which predicts a linear concentration profile
in the entire interval $[0,1]$. The discrepancy stems from part
(a) of the numerical scheme, which assumes that particles enter
the simulation interval exactly at $x=0$. However, $x=0$ is just
an imaginary interface. Had the simulation been run on the entire
line, particles would hop into the simulation across the imaginary
boundary at $x=0$ from the left, rather than exactly at the
boundary. This situation is familiar from renewal theory
\cite{Karlin}. The probability distribution of the distance an
entering particle covers, not given its previous location, is not
normal, but rather it is the residual of the normal distribution,
given by
\begin{equation}
f(x) = C \int_{-\infty}^0
\exp\left\{-\frac{(x-y)^2}{2\sigma^2}\right\}\,dy,
\end{equation}
where $\sigma^2 = \ds\frac{2\varepsilon \Delta t}{\gamma}$ and $C$
is determined by the normalization condition $\ds\int_0^\infty
f(x)\,dx = 1$. This gives
\begin{equation}
f(x) =
\sqrt{\frac{\pi}{2\sigma}}\,\mbox{erfc}\left(\frac{x}{\sqrt{2}\sigma}
\right).\label{eq:residual}
\end{equation}

\begin{figure}
\includegraphics[width=7in]{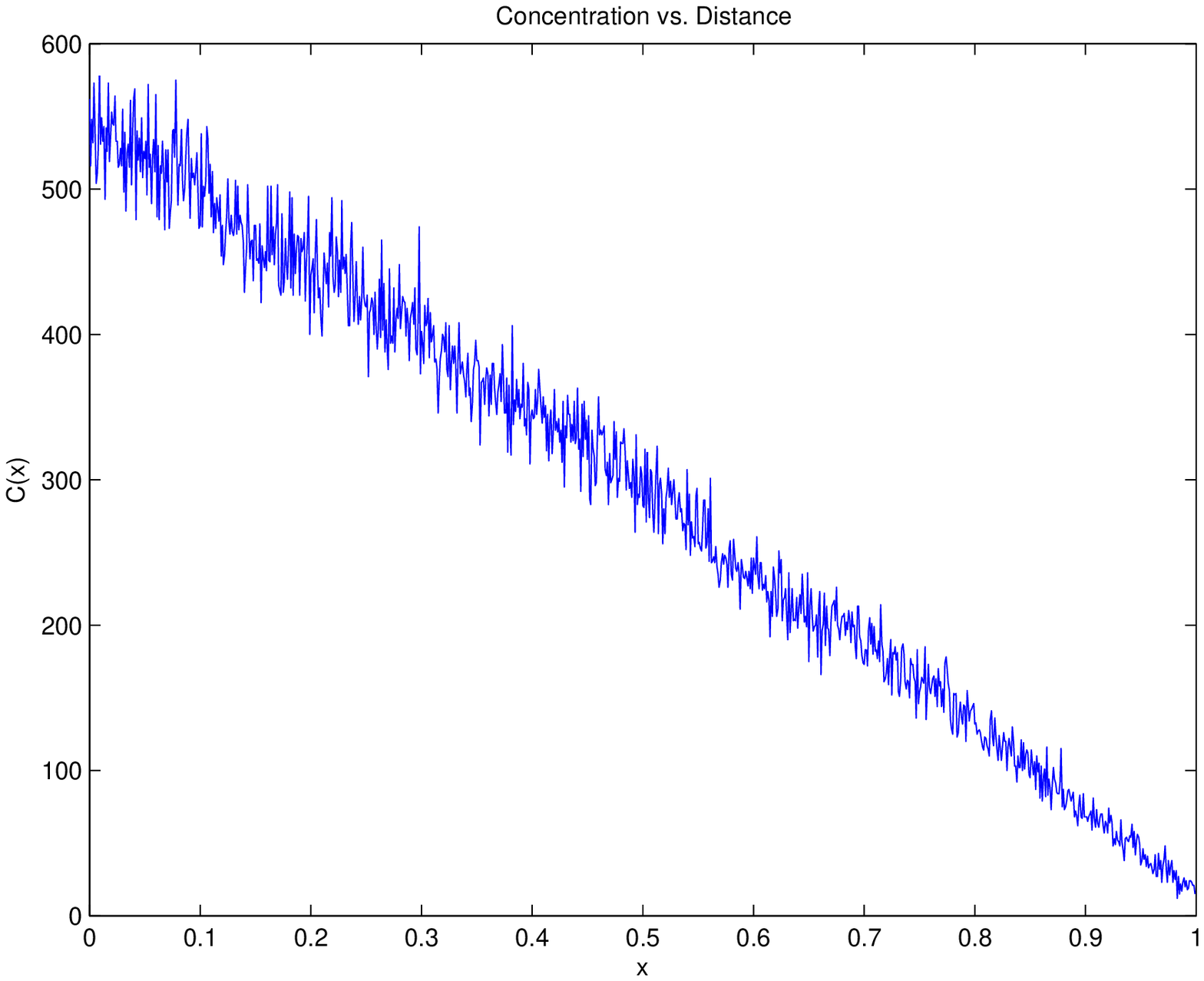}
\caption{The concentration profile of Brownian trajectories that
are initiated at $x=0$ with the residual of the normal
distribution, and terminated at either $x=0$ or $x=1$.
\label{f:residual}}
\end{figure}

Rerunning the simulation with the entrance pdf $f(x)$, we obtained
the expected linear concentration profile, without any depletion
layers (see Figure \ref{f:residual}). Injecting particles exactly
at the boundary makes their first leap into the simulation too
large, thus effectively decreasing the concentration profile near
the boundary.

\begin{figure}
\includegraphics[width=7in]{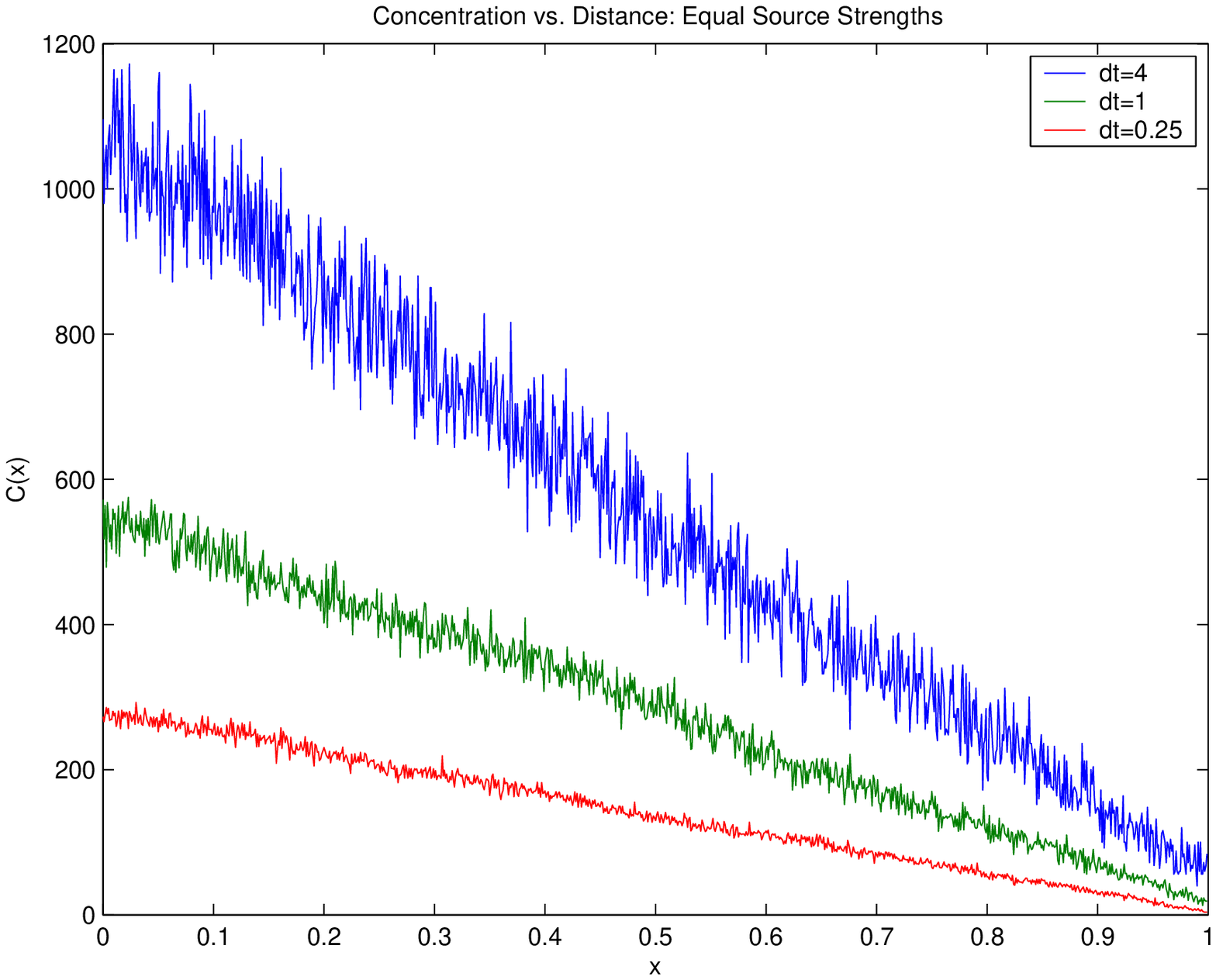}
\caption{The concentration profile of Brownian trajectories that
are initiated at $x=0$ and terminated at either $x=0$ or $x=1$.
Three different time steps $(\Delta t=4,1,0.25)$ were used, but
the injection rate of new particles remained constant. Refining
the time step decreases the concentration profile.
\label{f:equal_strengths}}
\end{figure}

\begin{figure}
\includegraphics[width=7in]{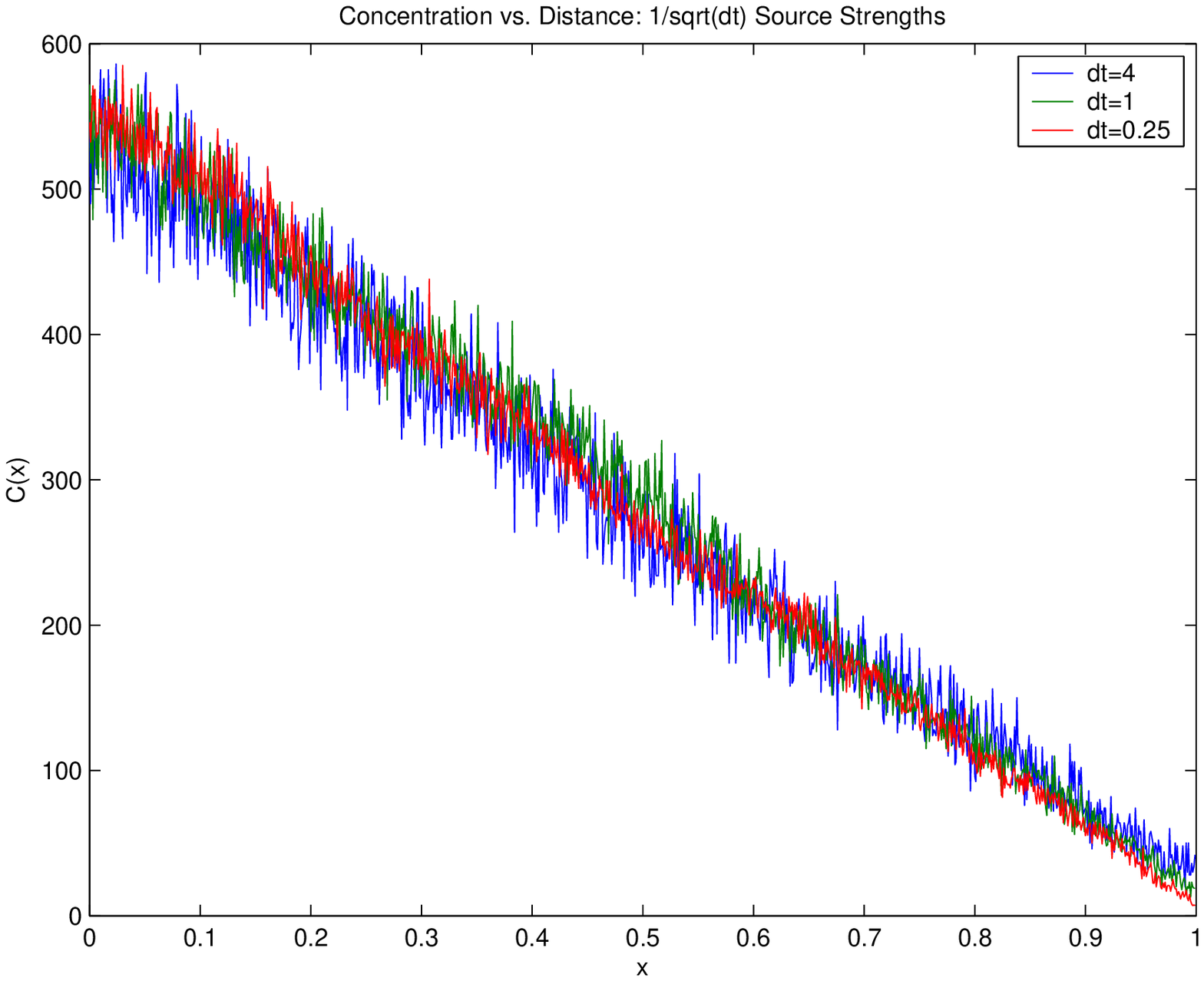}
\caption{The concentration profile of Brownian trajectories that
are initiated at $x=0$ and terminated at either $x=0$ or $x=1$.
Three different time steps $(\Delta t=4,1,0.25)$ are shown, and
the injection rate of new particles is proportional to
$1/\sqrt{\Delta t}$. Refining the time step does not alter the
concentration profile. \label{f:sqrt_strengths}}
\end{figure}

Next, we changed the time step $\Delta t$ of the simulation,
keeping the injection rate of new particles constant. The
population of trajectories inside the interval was depleted when
the time step was refined (see Figure \ref{f:equal_strengths}). A
well behaved numerical simulation is expected to converge as the
time step is refined, rather than to result in different profiles.
This shortcoming of refining the time step is remedied by
replacing the constant rate sources with time-step-dependent
sources, as predicted by eqs.(\ref{JLRexp})-(\ref{JRLexp}). Figure
\ref{f:sqrt_strengths} describes the concentration profiles for
three different values of $\Delta t$ and source strengths that are
proportional to $1/\sqrt{\Delta t}$. The concentration profiles
now converge when $\Delta t \to 0$. The key to this remedy is the
calculation of the UF in diffusion.

\section{Summary and discussion}
\label{sec:summary} Both Einstein \cite{Einstein} and Smoluchowski
\cite{Smoluchowski06} (see also \cite{Chandra}) pointed out that BD
is a valid description of diffusion only at times that are not too
short. More specifically, the Brownian approximation to the Langevin
equation breaks down at times shorter than $1/\gamma$, the
relaxation time of the medium in which the particles diffuse.

In a BD simulation of a channel the dynamics in the channel region
may be much more complicated than the dynamics near the interface,
somewhere inside the continuum bath, sufficiently far from the
channel. Thus the net flux is unknown, while the boundary
concentration is known. It follow that the simulation should be
run with source strengths (\ref{JLRexp}), (\ref{JRLexp}),
 \beqq
J_{LR}\sim\sqrt{\frac\varepsilon{\pi\gamma\Delta
 t}}\,C_L+\frac12J_{\mbox{net}},\quad J_{RL}\sim\sqrt{\frac\varepsilon{\pi\gamma\Delta
 t}}\,C_R-\frac12J_{\mbox{net}}.
 \eeqq
However, $J_{\mbox{net}}$ is unknown, so neglecting it relative to
$\sqrt{\ds\frac\varepsilon{\pi\gamma\Delta
 t}}\,C_{L,R}$ will lead to steady state boundary concentrations that are close, but
not necessarily equal to $C_L$ and $C_R$. Thus a shooting
procedure has to be adopted to adjust the boundary fluxes so that
the output concentrations agree with $C_L$ and $C_R$, and then the
net flux can be readily found.

According to (\ref{JLRexp}) and (\ref{JRLexp}), the efflux and
influx remain finite at the boundaries, and agree with the fluxes
of LD, if the time step in the BD simulation is chosen to be
$\Delta t=\ds\frac2\gamma$ near the boundary. If the time step is
chosen differently, the fluxes remain finite, but the simulation
does not recover the UF of LD. At points away from the boundary,
where correct UFs do not have to be recovered, the simulation can
proceed in coarser time steps.

The above analysis can be generalized to higher dimensions. In
three dimensions the normal component of the UF vector at a point
$\x$ on a given smooth surface represents the number of
trajectories that cross the surface from one side to the other,
per unit area at $\x$ in unit time. Particles cross the interface
in one direction if their velocity satisfies $\mb{v}\cdot
\mb{n}(\mb{x})>0$, where $\mb{n}(\mb{x})$ is the unit normal
vector to the surface at $\mb{x}$, thus defining the domain of
integration for eq.(\ref{Js}).

The time course of injection of particles into a BD simulation can
be chosen with any inter-injection probability density, as long as
the mean time between injections is chosen so that the source
strength is as indicated in (\ref{JLRexp}) and (\ref{JRLexp}). For
example, these times can be chosen independently of each other,
without creating spurious boundary layers.
\begin{acknowledgments}
The authors thank Dr. Shela Aboud for pointing out the depletion
phenomenon in BD simulations, and the referee for suggesting the
running of a BD simulation. The authors were partially supported
by research grants from the Israel Science Foundation, US-Israel
Binational Science Foundation, and DARPA.
\end{acknowledgments}

\newpage
\begin{figure}
\includegraphics[width=7in]{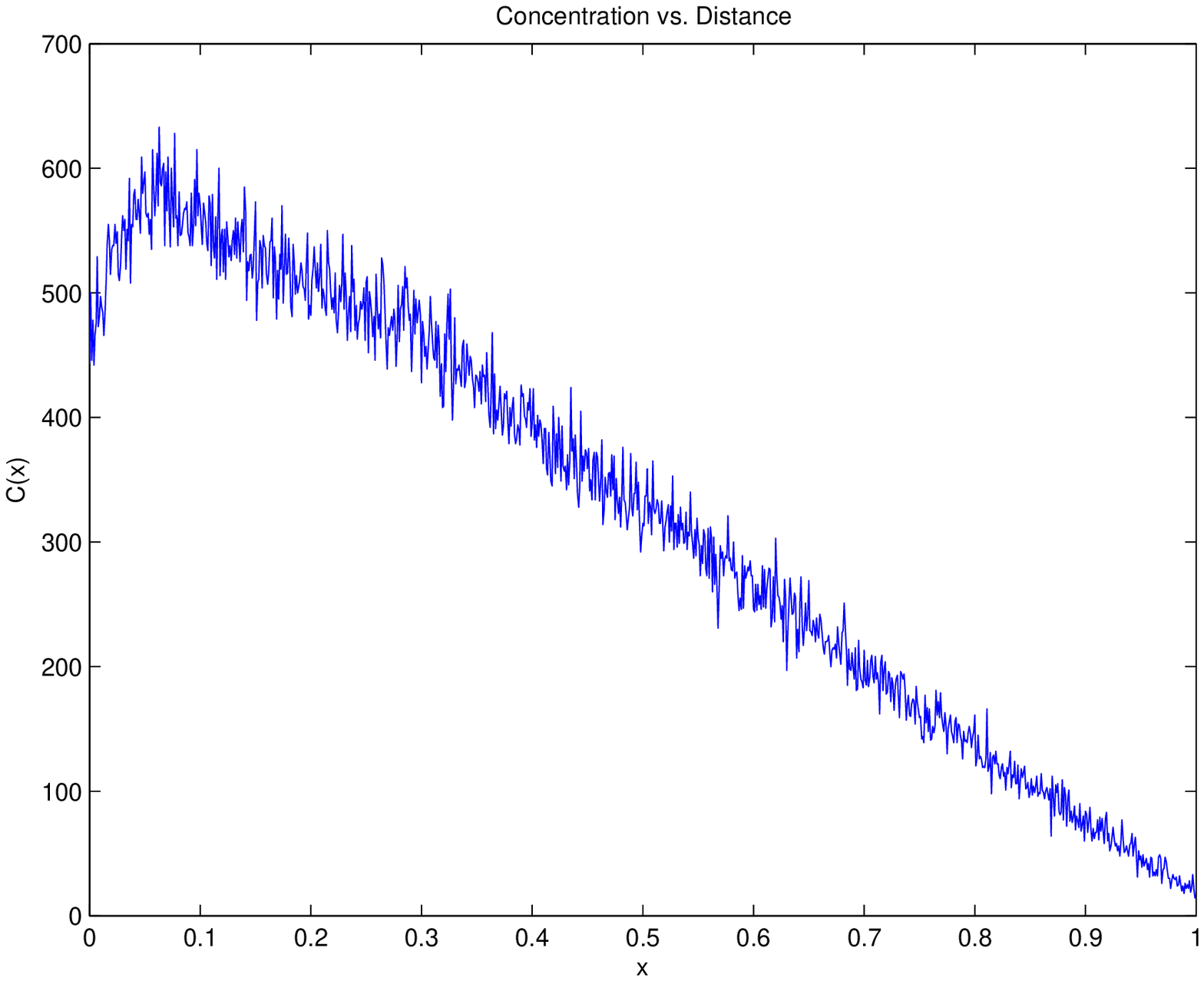}
\vspace{2cm} \\ A. Singer, PRE, Fig. 1
\end{figure}

\begin{figure}
\includegraphics[width=7in]{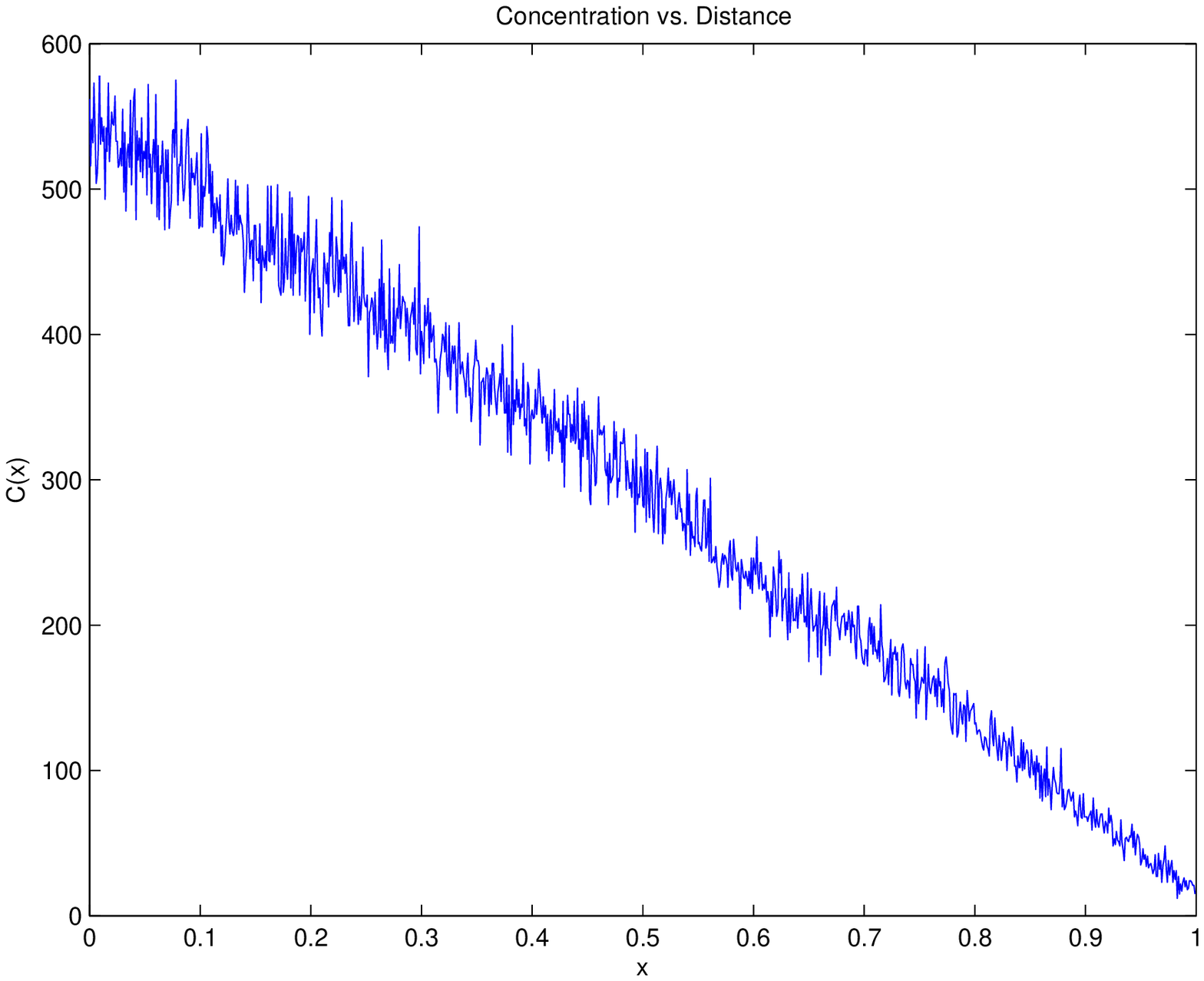}
\vspace{2cm} \\ A. Singer, PRE, Fig. 2
\end{figure}

\begin{figure}
\includegraphics[width=7in]{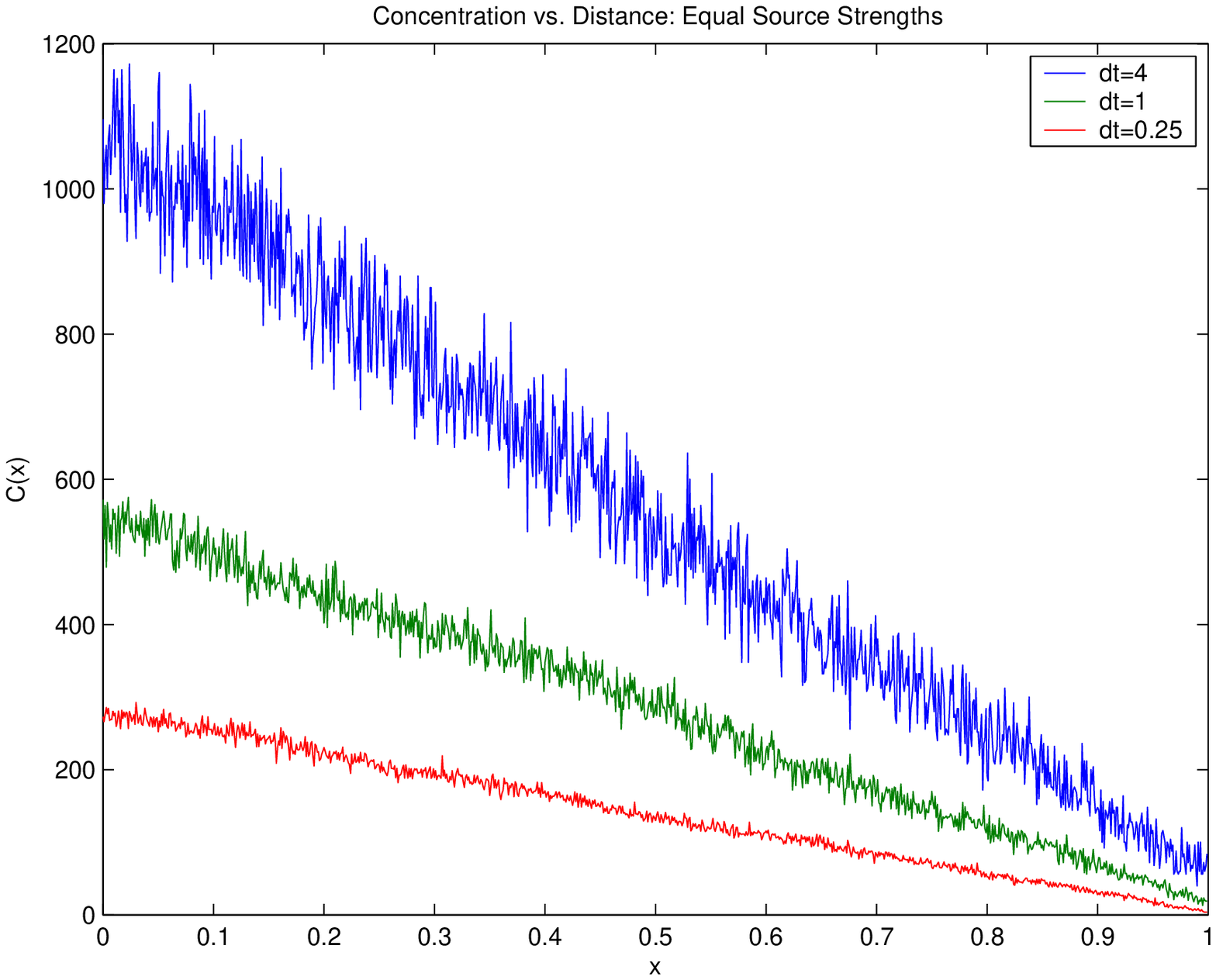}
\vspace{2cm} \\ A. Singer, PRE, Fig. 3
\end{figure}

\begin{figure}
\includegraphics[width=7in]{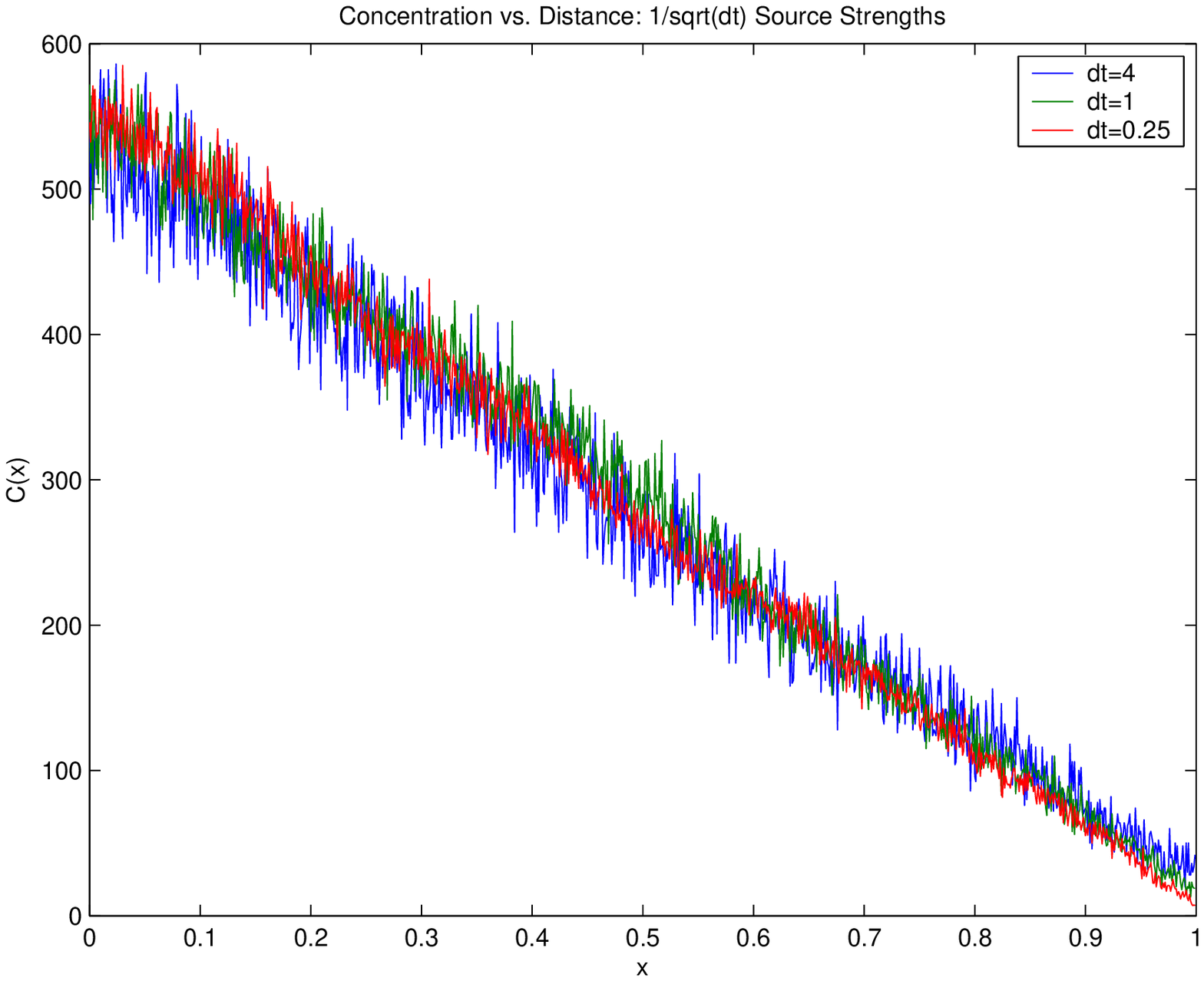}
\vspace{2cm} \\ A. Singer, PRE, Fig. 4
\end{figure}

\end{document}